
\magnification=\magstep1
\def\section#1{\bigskip\noindent{\bf #1}\medskip}
\def\subsection#1{\medskip\noindent #1\smallskip}

\centerline{STAR CLUSTERS\footnote*{to appear in in the Proceedings of
the conference {\sl Frontiers of Space and Ground Based Astronomy},
ESTEC, Noordwijk, The Netherlands, 10-14 May, 1993, eds. W. Wamsteker,
M.S. Longair and Y. Kondo, Kluwer, Dordrecht}
}
\bigskip
\centerline{Douglas C. Heggie}
\medskip
\centerline{University of Edinburgh,}
\centerline{Department of Mathematics and Statistics,}
\centerline{King's Buildings,}
\centerline{Edinburgh EH9 3JZ,}
\centerline{U.K.}
\bigskip
\bigskip

\centerline{Abstract}

This review concentrates almost entirely on globular star clusters.  It
emphasises the increasing realisation that few of the traditional
problems of star cluster astronomy can be studied in isolation: the
influence of the Galaxy affects dynamical evolution deep in the core,
and the spectrum of stellar masses; in turn the evolution of the core
determines the highest stellar densities, and the rate of encounters.
In this way external tidal effects indirectly influence the formation
and evolution of blue stragglers, binary pulsars, X-ray sources, etc.
More controversially, the stellar density appears to influence the
relative distribution of normal stars.  In the opposite sense, the
evolution of individual stars governs much of the early dynamics of a
globular cluster, and the existence of large numbers of primordial
binary stars has changed important details of our picture of the
dynamical evolution.  New computational tools which will become
available in the next few years will help dynamical theorists to address
these questions.

\vfill\eject

\section{1. Introduction}

The study of star clusters is pursued as vigorously today as it has
ever been. Their importance for astrophysics generally stems from
their role as relics of the formation of the Galaxy (including the
problem of their ages).  In addition they are an important test bed
for the theories of stellar evolution and stellar dynamics.  (No-one
would claim that understanding the dynamics of star clusters is one of
the most pressing problems of astronomy, but the processes which occur
in the star clusters, which are relatively easily observed, presumably
also occur in galactic nuclei.)

Another reason for the continuing high level of cluster studies is the
development of observational technique, which has extended studies from
the optical into the UV, IR, X-ray and radio wave bands.  Especially in
the latter two, unexpected new discoveries have opened up fresh theoretical
problems, while at the same time shedding light on areas that were
already under investigation for other reasons.

This review looks at globular star clusters from the {\sl dynamical}
point of view.  This bias is less serious than might be thought,
because it is gradually being realised that the main problems of star
clusters cannot be understood in isolation: for example, the dynamics
of the clusters depends, in part, on the rate at which stars lose mass
by internal evolution; while the dynamics controls the rate at which
stars interact, and hence (in all probability) the formation of the
cluster X-ray sources, millisecond pulsars, blue stragglers, etc.  The
idea that such aspects of cluster evolution are all interrelated is
not entirely new, but for many years most research proceeded as though
they could be understood independently of each other.  Now, each
specialty in cluster studies is finding that further progress depends
on viewing each star cluster as a kind of ecosystem of interrelated
species.  Therefore, when we look a little deeply at the clusters from
the dynamical point of view, we see the same objects as those studied
by the X-ray astronomer or the observer interested in colour-magnitude
diagrams, and many of the same problems.

\section{2. External influences}

For many years dynamicists thought that the study of globular clusters
was fundamentally the same as that of isolated stellar systems.  It
was known that clusters were tidally limited, in a way that was
successfully described by King models (King 1966), but the influence
of the tide on the behaviour deep in the core was ignored until the
work of Spitzer \& Chevalier (1973), who showed that the rate of core
collapse was greatly affected by tidal effects.  This prediction has
since been amply confirmed by surveys of the surface density profiles
of observed clusters, which show that clusters with collapsed cores
are prevalent at small galactocentric distances (Chernoff \&
Djorgovski 1989), where the tidal effects are strongest.

(Incidentally, the identification of post-collapse clusters, starting
with M15, is one of the significant triumphs of the last decade or
two.  It brings to a conclusion a story started with the classical
theoretical study of H\'enon (1961).  Even though the core of M15 now
turns out to be resolvable (Lauer {\sl et al} 1991; but cf. Yanny {\sl
et al} 1993), it is still best understood as a post-collapse cluster
(Grabhorn {\sl et al} 1992).)

While the Galactic tide controls the extent and influences the core
dynamics of the existing clusters, it is also one of the main
destructive processes.  At present clusters are destroyed by this and
other processes at a rate of about 5 per Gyr (Hut \& Djorgovski 1992).
Consideration of the joint action of tidal processes and internal
evolution helps to explain the observed relatively narrow distribution
of cluster masses at the present (Fall \& Rees 1977).  On the other hand
studies of the median cluster mass in other (extragalactic) cluster
systems suggests that this is almost independent of the strength of
tidal effects (van den Bergh 1993a), and so the role of tides may not be
decisive.

Where the tide continues to influence the properties of existing
clusters is in the spectrum of stellar masses.  Internal processes drive
the less massive stars to larger radii within the cluster, where they
are relatively more easily stripped by the tide (cf.  Chernoff \&
Weinberg 1990).  Now that star counts in a number of clusters extend to
sufficiently faint magnitudes, it can be seen that the present-day mass
function is flattest (most depleted in low masses) in clusters which are
most subject to tidal disruption (Capaccioli {\sl et al} 1993).

\section{3. Stellar Influences on Cluster Dynamics}

\subsection{3.1 Mass Loss}

We have seen that external influences have a large effect on the size,
stellar population and internal dynamics of star clusters.  Now we turn
to the ways in which the stellar population itself affects the dynamics.
One of the most important is the mass lost in the evolution of the more
massive stars.  Obvious though this effect is, it was almost entirely
ignored in dynamical studies for many years, with the notable exception of
the work of Angeletti \& Giannone (1977).  Recent work on this mechanism
stems from a paper by Applegate (1986).  Mass loss through stellar
evolution leads
to an increase in the radius of a cluster.   Indeed the mean core
radii of the older clusters in the LMC exceed those of the younger
objects, but the magnitude of the effect is not yet in satisfactory
quantitative agreement with dynamical models (Elson 1992).

Mass loss through stellar evolution is a serious disruptive process
especially when the initial mass function is relatively flat (Chernoff
\& Weinberg 1990).  To some extent this may be an artefact of the
dynamical method they used (Fokker-Planck models),
as $N$-body models indicate that clusters may survive
a period of severe mass loss after a period of dynamical readjustment
(Fig.1), even though the Fokker-Planck model suggests that clusters with
identical parameters should disrupt.
The fact that a cluster may well survive even with a
relatively shallow mass spectrum helps to understand how the clusters
may still possess enough neutron stars (born from the evolution of very
high-mass stars) to explain the observed numbers of X-ray sources
and pulsars.

One should add that the initial parameters of the models shown in Fig.1
and discussed in Chernoff \& Weinberg (1990) are quite restrictive: in
addition to the mass spectrum, the survival of a cluster depends on its
initial size relative to the tidal radius, a parameter that has been
relatively little explored (but see Phinney 1992). This initial
condition probably also leaves an imprint on the anisotropy of the
stellar velocity distribution.  Anisotropy is manifest in proper motion
studies of clusters such as M13 (Cudworth {\sl et al} 1985), whereas
$N$-body models (Giersz \& Heggie 1993) show little anisotropy if the
initial radius of the system is too large.

\vfill\eject
\subsection{3.2 Binaries}

The second main influence of the stellar content of a cluster on its
dynamics is the presence of binary stars.  As with the role of stellar
mass loss, their effects were almost completely disregarded until
recently, with a few exceptions (Hills 1975, Spitzer \&
Mathieu 1980).  What has transformed the picture is the observational
discovery of binaries of various types, using an increasing variety of
techniques (see Hut {\sl et al} 1992 for a review and many further
references).  Their effects on the dynamics are two-fold. First, they
limit the depth of core collapse, since encounters between binaries
generate kinetic energy more efficiently than other processes (Fig.2).
Second, their interactions can power much of the post-collapse evolution
of a cluster.  As they interact they are destroyed, like a non-renewable
source of energy, but in tidally limited models the reserves may be
sufficient to sustain a cluster for its entire lifetime.

The discovery of primordial binaries has given us a picture of
dynamical evolution which is in some ways simpler than those
previously available.  One of the central questions was the dominant
mechanism for generating energy.  The most promising answer was the
evolution of binaries in tidal two-body encounters (Statler {\sl et
al} 1985, Stod\'o\l kiewicz 1985), but this was complicated by their
small semi-major axis: in any three-body interaction at least two of
the stars are bound to collide (Hut \& Inagaki 1985, Cleary \&
Monaghan 1990).  Dynamically the most effective primordial binaries
are those with much larger semi-major axes, for which this problem is
only now being explored (McMillan 1993).  As we shall see, an
understanding of stellar collisions is still of great importance, but
for different reasons.

\section{4. Dynamical influences on the Stellar Content}

In discussing the role of tidal processes on the stellar mass function
we have already mentioned one way in which dynamics influences the kinds
of stars found in globular clusters.  Another, which has been known for
a long time and has been the impetus for some observational studies
(e.g.  Richer \& Fahlman 1989), is mass segregation.

Mass segregation might be expected to give rise to a colour gradient,
but the question of colour gradients in globular star clusters has an
unhappy history, because the centre of a cluster may be assigned to what
is nothing more than a chance concentration of giants, and will
therefore appear red.  In recent years, however, it has been found
(Djorgovski {\sl et al} 1991; see also Cederbloom {\sl et al} 1992) that
most clusters exhibit flat colour profiles.  Most interestingly, the
exceptions are mainly those with post-core collapse profiles, which tend
to be bluer at the centre.  This is one of the clearest and most
puzzling observational indications that dynamics influences the stellar
content of globular star clusters.  Interestingly, it was foreshadowed by
early IUE observations (Dupree {\sl et al} 1979) which indicated that
the core radius was smaller at shorter wavelengths (cf. also Djorgovski
\& Piotto 1992).

What feature of the stellar distribution is responsible for the colour
gradient? Recent work (Djorgovski {\sl et al} 1991) has confirmed for
several clusters an older finding for M15 (Auriere \& Cordoni 1981) that
red giants are relatively depleted in the central regions of
post-collapse clusters.  By contrast the distribution of stars on and
near the horizontal branch is quite strongly enhanced in dense cores.

These issues have important repercussions in the study of
colour-magnitude diagrams.  The tip of the red giant branch is truncated
in denser stellar environments (Djorgovski \& Piotto 1993).  Even more
dramatic are the effects on the horizontal branch (e.g. Bailyn {\sl et
al} 1992), and this has rather wide implications in a variety of cluster
studies.  It is well known that the primary parameter which determines
the morphology of the horizontal branch is the metal abundance (cf. Lee
1991 for a recent discussion), but it has been known for many years that
at intermediate abundances the morphology may be very different in two
clusters which have the same abundance; in other words some ``second
parameter" influences its morphology. To many experts the best candidate
is the age of the cluster, but it now appears that the stellar density
is another plausible candidate (Buonanno 1993).  At any rate, it is no
longer very meaningful to refer to {\sl the} colour-magnitude diagram of
a globular cluster: for some clusters, important details vary from one
part of the cluster to another.

Mass segregation appears to work in the wrong direction to account for
these observations, and it seems likely that an explanation will involve
an understanding of physical stellar encounters.  Fortunately, this
difficult problem has been under study now for some time (e.g. Davies
{\sl et al} 1991, Benz \& Hills 1992, Rasio 1993), motivated partly by
attempts to understand the population of blue stragglers.  Now that
high-resolution studies even of relatively dense cluster cores are
becoming possible (Yanny {\sl et al} 1993, Ferraro 1993) these are being
found in increasing numbers in several clusters.  They may well have
more than one origin (e.g. Bailyn 1992), but mergers resulting from
stellar collisions is likely to dominate in some clusters.

Finally we turn to the high-energy astrophysicist's view of globular
clusters (cf. Bailyn 1991 for a review).  The variable X-ray sources
discovered in the 1970s are binary stars in which one component is a
neutron star, and the more recent discovery of millisecond pulsars, some
of which are in binaries, has further opened a new window into dynamical
processes occurring in star clusters.  Two considerable uncertainties in
explaining the origin of these objects are, first, the retention of
neutron stars in the shallow potential well of a typical globular
cluster, and, second, the probability that the birth of a neutron star
within a binary may well disrupt the binary.  Hills (1977) suggested that
the solution to the latter problem  lay with exchange reactions,
in which a single neutron star is likely to displace one component of a
binary in an encounter, because of its relatively high mass.  A recent
encounter between a binary pulsar and a single star in M15 is the most
plausible explanation of both its eccentricity and its large distance
{}from the core of the cluster (Phinney \& Sigurdsson 1991).

\section{5. Some Highlights of Cluster Research} The foregoing
sections of this review span the links between the environment, the
internal dynamics and the stellar content of globular star clusters.
They also show how observations at various wavelengths are used to
help provide an integrated picture of these aspects; for example,
radio observations of millisecond pulsars, optical photometry for
colour-magnitude diagrams, ultra-violet surface brightness profiles,
X-ray studies, and optical observations from space.  On the other hand
the previous sections fail to touch on some of the central problems
which have engaged much attention in recent years.  The aim of the
present section is to redress the balance a little by underscoring
some important recent advances and problems.

Foremost among these must be the realisation that even the Galactic
system of globular clusters is relatively differentiated by kinematics,
Galactic distribution and composition (see Zinn 1990, 1991 for reviews).
At the very least the system can be divided into two components, and
there is even an extreme view that there is no real distinction between
the youngest globular clusters and the oldest open clusters.

The implications of these findings for theories of cluster formation and
galactic evolution are less clear, partly because the relative dating of
the two main classes of Galactic globular clusters has not yet been
determined. It is even possible that the diversity in several cluster
properties reflects successive mergers rather than changing
circumstances of formation within the Galaxy (e.g.  van den Bergh
1993b).

Motivated largely by the cosmological problem of the age of the
universe, substantial effort has been applied to the determination of
cluster ages.  While it is now clear that there is a considerable spread
in ages, the main thing that has been learned about absolute ages is the
existence of several factors which make this question difficult to answer.

Much research has been devoted to the investigation of elemental
abundances within individual clusters (see Norris 1988 for a review).
While some variations (within an individual cluster) are associated with
stellar evolution, in all clusters except two ($\omega$Cen and M22) the
variations in primordial abundances are small, and in some cases the
variations in the abundance of iron are less than $0.15$dex.

Of importance for the modelling and dynamics of star clusters (see, for
example, Meylan \& Pryor 1993) is the accurate and increasingly
extensive measurement of individual stellar radial velocities.
Complementary information has come from the ever improving study of
proper motions, with the result that for several clusters we now have
three-dimensional information on both space and internal motions (e.g.
Cudworth 1993).

Studies of globular clusters have gradually extended to include
extragalactic cluster systems. Confined at first to the study of
integrated properties, technical advances now make it possible to study
colour-magnitude diagrams and structural properties of clusters in
nearby galaxies (e.g.  Christian \& Heasley 1991, Cacciari {\sl et al}
1993 and references therein).  Finally, this brief review does no
justice at all to the nearest cluster system -- the open clusters of the
Galaxy, despite their importance for such central problems as the
distance scale, Galactic structure, and star formation.

\section{6. The Next Few Years}
In this concluding section we revert largely to the theoretical aspects
which were discussed in \S\S2-4.  We shall not even discuss the
scientific problems specifically, but consider instead some of the tools
which are used by theorists, and how they may be expected to develop in
the foreseeable future.

The tool which many theorists would like to apply to these problems is
$N$-body simulation, but it is exceptionally time-consuming.  Great
advances have been made in the last 30 years, partly because of
improvements in general-purpose hardware, and partly because of the
development of a small number of highly tuned and efficient codes which
have been specially devised with these problems in mind (Aarseth 1985).
These tools are sufficiently well developed to provide numerical
simulations of systems with as many as 10000 particles (Aarseth \&
Spurzem 1993; this paper, Fig.3), though not yet routinely.  This is
more than enough, however, for very detailed and realistic models of
open star clusters, and yet little has been done on this problem with
these methods since the work of Terlevich (1987).

At the present time the capabilities of $N$-body methods are
undergoing rapid transformation, as special purpose {\sl hardware} is
currently under development (Mak\-ino {\sl et al} 1992).  Within a
year or two it should be possible for the first time to model a
cluster containing, say, 50000 stars.  This is still small relative to
a typical globular cluster, but it goes a long way to closing the
number-gap which at present must be bridged with more or less
simplified theories.  What makes these developments still more
relevant to the kinds of problems discussed in \S\S2-4 is that they
should be able to include fairly realistic treatments of close stellar
interactions, by means of smooth particle hydrodynamics, for example
(cf.  Rasio 1993).  In this way one might hope for a better
theoretical understanding of such questions as the expected numbers of
blue stragglers, etc.

Continuing developments should eventually enable us to provide realistic
models of individual star clusters.  Indeed the most serious foreseeable
difficulty is in correctly modelling the astrophysical aspects of such
$N$-body models, e.g.  the internal stellar evolution of binary stars.
But even when such models become possible, some time must elapse before
they become routine.  Until this happens it is likely that much can be
learned by rather more simplified techniques.  One which deserves to be
revived is the Monte Carlo method originally developed by H\'enon (1973)
and brought to an amazing level of realism by Stod\'o\l kiewicz (1985).
It was somewhat eclipsed by the finite-difference method of Cohn (1979),
but the latter becomes relatively rather expensive when it is made more
realistic, unlike the Monte Carlo method.  Until direct $N$-body
simulation routinely achieves the required capability, the Monte Carlo
method may well be the most effective tool for bridging the gap between
the largest $N$-body simulations and the globular clusters.

\bigskip\noindent
References

\leftskip=0.3truein\parindent=-0.3truein

Aarseth S.J., 1966, MNRAS, 132, 35

Aarseth S.J., 1968, Bull.Astron., 3, 105

Aarseth S.J., 1973, Vistas Astron., 15, 13

Aarseth S.J., 1985, in Brackbill J.U., Cohen B.I., eds,  Multiple Time
Scales.  Academic Press, New York, p. 377

Aarseth S.J., Heggie D.C., 1992, MNRAS, 257, 513

Aarseth S.J., Heggie D.C., 1993, in Djorgovski S., Meylan G., eds,
Structure and Dynamics of Globular Clusters, ASP Conference Series,
Vol. 50.  ASP, San Francisco, in press

Aarseth S.J., Spurzem R., 1993, in preparation

Angeletti L., Giannone P., 1977, A\&A, 58, 363

Applegate J.H., 1986, ApJ, 301, 132

Auriere M.,  Cordoni J.-P., 1981, A\&A,
100, 307

Bailyn C.D., 1991, in Janes K., ed, The Formation and Evolution of Star
Clusters, ASP Conf. Ser. Vol.13. ASP, San Francisco, p.307

Bailyn C.D., 1992, ApJ, 392, 519

Bailyn C.D., Sarajedini A., Cohn H., Lugger P.M., Grindlay J.E., 1992,
AJ, 103, 1564

Benz W.,   Hills J.G., 1992, ApJ, 389, 546

Buonanno R., 1993, in Brodie J., Smith G., eds, The
Globular Cluster - Galaxy Connection, Proc. 11th Santa Cruz Summer
School, ASP Conf. Ser. ASP, San Francisco, in press

Cacciari C., Battistini P., Bendinelli O., Bonoli F., Buonanno
R.,  Djorgovski S.,  Federici L.,  Ferraro F., Fusi Pecci F., King
I.R.,  Parmeggiani G., Walterbos R.,   Zavatti F., 1992, in Science with
the Hubble Space Telescope, Proc.
ST-ECF/STScI Workshop. ESA, in press

Capaccioli M., Piotto G.,   Stiavelli M., 1993, MNRAS, 261, 819

Cederbloom S.E., Moss M.J., Cohn H.N., Lugger P.M., Bailyn C.D.,
Grindlay J.E.,  McClure R.D., 1992, AJ, 103, 480

Chernoff D.F., Djorgovski S., 1989, ApJ, 339, 904

Chernoff D.F., Weinberg M.D., 1990, ApJ, 351, 121

Christian C.A., Heasley J.N., 1991, AJ, 101, 848

Cleary P.W., Monaghan J.J., 1990, ApJ,  349, 150

Cohn H., 1979, ApJ, 234, 1036

Cudworth K., 1993, in  Djorgovski S.,  Meylan G., eds, Structure and
Dynamics of Globular Clusters, ASP Conf. Ser., Vol. 50. ASP,
San Francisco, in press

Cudworth K., Lin D.N.C., Oh K.-S., 1985, in Goodman J., Hut P., eds,
Dynamics of Star Clusters, IAU Symp. 113, Reidel, Dordrecht, p.65

Davies M.B., Benz W.,  Hills J.G., 1991, ApJ, 381, 449

Djorgovski S., Piotto G., 1992, AJ, 104, 2112

Djorgovski S., Piotto G., 1993, in Brodie J., Smith G., eds, The
Globular Cluster - Galaxy Connection, Proc. 11th Santa Cruz Summer
School, ASP Conf. Ser. ASP, San Francisco, in press

Djorgovski S., Piotto G., Phinney E.S.,  Chernoff D.F., 1991, ApJ, L41

Dupree A.K., Hartmann L., Black J.H., Davis R.J., Matilsky T.A.,
Raymond J.C.,  Gursky H., 1979, ApJ, 230, L89

Elson R.A.W., 1992, MNRAS, 256, 515

Fall S.M.,  Rees M.J., 1977, MNRAS, 181, 37P.

Ferraro F.R., 1993, this volume

Giersz M., Heggie D.C., 1993, in preparation

Grabhorn R.P., Cohn H.N., Lugger P.M., Murphy B.W., 1992,
ApJ, 392, 86

H\'enon M., 1961, Annales d'Astrophysique, 24, 369.

H\'enon M., 1973, in Martinet L., Mayor M., eds, Dynamical Structure and
Evolution of Stellar Systems, SAAS-FEE, 3rd Advanced Course.  Geneva
Observatory, p.224

Hills J.G., 1975, AJ, 80, 1075

Hills J.G., 1977, AJ, 82, 626

Hut P.,  Djorgovski S., 1992, Nature, 359, 806

Hut P., Inagaki S., 1985, ApJ, 298, 502

Hut P., McMillan S.L.W., Goodman J., Mateo M., Phinney E.S.,
Pryor C., Richer H.B., Verbunt F.,  Weinberg M., 1992, PASP, 104, 981

Inagaki S., 1986, PASJ, 38, 853

King I.R., 1966, AJ, 71, 64

Lauer T.R., Holtzman J.A., Faber S.M., Baum W.A., Currie D.G.,
Ewald S.P., Groth E.J., Hester J.J., Kelsall T., Light R.M., Lynds
C.R., O'Neill E.J., Jr., Schneider D.P., Shaya E.J., Westphal J.A.,
1991, ApJ, 369, L45.

Lee Y.-W., 1991, in Janes K., ed, The Formation and Evolution of Star
Clusters, ASP Conf. Ser. Vol.13. ASP, San Francisco, p.205

Makino J., Kokubo E.,  Taiji M., 1992, ``HARP: A Special-Purpose Computer
for N-Body Problem", preprint

McMillan S.L.W., 1993, in Djorgovski S., Meylan G., eds, Structure and
Dynamics of Globular Clusters, ASP Conference Series, Vol. 50.  ASP,
San Francisco, in press

Meylan G., Pryor C., 1993,  in  Djorgovski S.,  Meylan G., eds,
Structure and Dynamics of Globular Clusters, ASP Conf. Ser., Vol. 50. ASP,
San Francisco, in press

Norris J.E., 1988, in
Grindlay J.E., Philip A.G.D., eds, The Harlow Shapley
Symposium on Globular Cluster Systems in Galaxies, IAU Symp. 126.
Kluwer, Dordrecht, p.93

Phinney E.S.,  Sigurdsson S., 1991, Nature, 349, 220

Phinney E.S., 1992, Phil. Trans. R. Soc. Lond. A, 341, 39

Rasio F., 1993, ``Hydrodynamic Stellar Interactions in Dense Star
Clusters", Proc 2nd Hubble Symp., ASP Conf. Ser. ASP, San Francisco,
to appear

Richer H.B.,  Fahlman G.G., 1989, ApJ, 339, 178

Spitzer L., Jr., Chevalier R.A., 1973, ApJ, 183, 565.

Spitzer L., Jr.,  Mathieu R.D., 1980, ApJ, 241, 618

Statler T.S., Ostriker J.P.,  Cohn H.N., 1985,  ApJ, 316, 626

Stod\'o\l kiewicz J.S., 1985,  in Goodman J., Hut P., eds, Dynamics
of Star Clusters, IAU Symp. 113. Reidel, Dordrecht, p.361

Terlevich E., 1980, in Hesser J.E., Star Clusters, IAU Symp. 85. Reidel
Dordrecht, p.165

Terlevich E., 1987, MNRAS,  224, 193

van den Bergh S., 1993a, in  Djorgovski S.,  Meylan G., eds, Structure
and Dynamics of Globular Clusters, ASP Conf. Ser., Vol. 50. ASP,
San Francisco, in press

van den Bergh S., 1993b, MNRAS, 262, 588

von Hoerner S., 1960, Z.Ap., 50, 184

von Hoerner S., 1963, Z.Ap., 57, 47

Yanny B., Guhathakurta P., Schneider D.P., Bahcall J.N., 1993, ``Hubble
Space Telescope Observations of the Center of Globular Cluster M15",
PASP, Proc 2nd Hubble Symp., to appear

Zinn R., 1990, JRASC, 84, 89

Zinn R., 1991, in Janes K., ed, The Formation and Evolution of Star
Clusters, ASP Conf. Ser. Vol.13. ASP, San Francisco, p.532
\vfill\eject
\leftskip=0pt
\parindent=0pt
\parskip=\medskipamount

\bigskip\noindent
Figure Captions

Fig.1 Evolution of core radius in a large ensemble of 500-body systems
(after Giersz \& Heggie 1993).  The initial conditions were King
models with the stated scaled central potential $W_0$, and with a mass
spectrum $f(m)\propto m^{-x}$.  Tidal parameters and total mass (see
title) have been chosen appropriate to the LMC. According to
Fokker-Planck models of Weinberg (cf. Elson 1992) a cluster with
$x=0.5$ and $W_0=7$ should disrupt after about $3\times 10^7$yr.

Fig.2 Evolution of  core radius in $N$-body models with and without
primordial binaries.  The initial conditions are described fully in
Heggie \& Aarseth (1992), where they are referred to as Models S and I.
With primordial binaries the depth of core collapse is much shallower,
and the effect would be greater in larger systems than in these
2500-body systems.

Fig.3 Advances in  collisional $N$-body simulations.  Only large
models extending a long way into core collapse, or beyond, are included.
For the most part the date is that of publication, except for the
last points, where dates are provisional.
Sources can be traced from the name and date, except
for HARP (Makino {\sl et al} 1992) for which data are conservative
estimates only.

\bye